# Superconducting junction of a single-crystalline Au nanowire for an ideal Josephson device

Minkyung Jung,[†,1] Hyunho Noh,[†] Yong-Joo Doh,[§] Woon Song,[†] Yonuk Chong,[†] Mahn-Soo Choi,[§§] Youngdong Yoo,[‖] Kwanyong Seo,[‖] Nam Kim,[†] Byung-Chill Woo,[†] Bongsoo Kim,[‖,*] and Jinhee Kim[†,*]

[†] *Korea Research Institute of Standards and Science, Daejeon 305-600, Korea*

[§] *Department of Display and Semiconductor Physics, Sejong Campus, Korea University Sejong Campus, Chungnam 339-700, Korea*

[§§] *Department of Physics, Korea University, Seoul 136-713, Korea*

[‖] *Department of Chemistry, KAIST, Daejeon 305-701, Korea*

[1]Present address: Department of Physics, Princeton University, Princeton, NJ 08544, USA

**We report on the fabrication and measurements of a superconducting junction of a single-crystalline Au nanowire, connected to Al electrodes. Current-Voltage characteristic curve shows clear supercurrent branch below the superconducting transition temperature of Al and quantized voltage plateaus on application of microwave radiation, as expected from Josephson relations. Highly transparent (0.95) contacts very close to an ideal limit of 1 are formed at the interface between the normal metal (Au) and the superconductor (Al). The very high transparency is ascribed to the single crystallinity of a Au nanowire and the formation of an oxide-free contact between Au and Al. The sub-gap structures of the differential conductance are well explained by coherent multiple Andreev reflections (MAR), the hallmark of mesoscopic Josephson junctions. These observations demonstrate that single crystalline Au nanowires can be employed to develop novel quantum devices utilizing coherent electrical transport.**

When a non-superconducting electron system, referred to as a "normal system", is in contact with a superconductor, the Cooper pairs in the superconductor penetrate the normal system, thereby inducing a superconducting order parameter in it. This is the well-known proximity effect.[1-9] The most commonly used method to investigate the proximity effect is transport measurement through superconductor - normal system - superconductor (SNS) junctions. The earliest studies using this approach employed samples in the semi-classical regime, where phase coherence of Cooper pairs in the normal system was ignored.[1-3] Advances in micro-fabrication technology have enabled realization of a SNS junction in the mesoscopic regime, where the phase coherence can be maintained throughout the normal system. SNS junctions with semiconducting two-dimensional electron gas (2DEG),[10, 11] atomic point contacts,[12] semiconducting nanowires,[13, 14, 34] carbon nanotubes,[15, 16] and, more recently, graphene[17] as a normal system have been successfully fabricated and studied. In this mesoscopic regime, the measured physical effects exhibit ubiquitous influence of phase-coherent transport of Cooper pairs penetrating the normal system.

The theory of proximity effect was initially investigated by de Gennes.[1, 2] Various theories have since been developed to incorporate the phase-coherent transport of Cooper pairs through SNS junctions, both in the diffusive limit[3-5] and ballistic limit.[6-9] Underlying these theories is the picture of multiple Andreev reflections (MARs). An electron in a normal system slightly above the Fermi level is reflected back as a hole at one super conductor - normal system (NS) interface leaving a Cooper in the superconductor. The hole is reflected again as an electron at the other NS interface, and these processes are repeated coherently, resulting in a resonance-like behaviour. The MAR is the hallmark of phase-coherent transport of Cooper pairs in mesoscopic Josephson junctions.

Even though in his original theory of the proximity effect[1] de Gennes had in mind "metal" as the normal system, MARs and coherent Cooper pair transport have been observed mostly in SNS junctions with semiconductor or semimetal as a normal system. The observation of MARs in an all-metallic system is presumably hindered by the high critical current of the junctions, which results in severe Joule heating, and also by the relatively low transparency of the NS interface.[18] Herein, we report on the very unique features of mesoscopic Josephson junctions comprised of single crystalline Au NWs connected to superconducting Al electrodes, an ideal system that is much closer in spirit to the model considered by de Gennes. The Au NWs, grown by the vapor transport method,[19] are single-crystalline with no defects and do not form a native oxide layer on their surface, thereby enabling us to form oxide-free and highly transparent contact between Al and Au of the NS interface. We observed supercurrent flow through the Au NW and multiple peaks of dynamic conductance caused by resonant MARs. Single-crystalline Au NWs hold promise as a facile route to realize various quantum devices with π-junctions[20] and Andreev qubits.[21]

Figure 1(a) displays a scanning electron microscopy (SEM) image of Au NWs and a high-resolution transmission-electron microscopy (TEM) image in the inset. Details of the NW growth and device fabrication have been reported elsewhere.[19, 22] A representative SEM image and a schematic view of the device are shown in Fig. 1(b) and the inset, respectively. The width ($w$) of the Au NW and the distance ($L$) between two superconducting electrodes are found to be $w = 80 - 125$ nm and $L = 280 - 740$ nm, respectively. The normal-state resistance ($R_n$) of the junctions ranges from 4 to 10 Ω, and thus the upper limit of the Au NW resistivity is found to be $\rho_{Au} = 14 \pm 8$ μΩcm at $T = 2$ K.

As the temperature decreases, the junction resistance drops sharply at around $T = 1.3$ K, which corresponds to the superconducting transition temperature ($T_c$) of the Al electrodes. Below $T_c$, the resistance diminishes to zero near $T = 1.2$ K due to supercurrent flow through the NW. Figure 2(a) shows typical current–voltage ($I$–$V$) characteristics at a temperature of $T = 260$ mK. In the figure, a clear supercurrent branch and also a resistive quasiparticle branch can be observed. When the bias current $I$ increased from zero, a sudden switching from supercurrent to the resistive branch occurs at a critical current ($I_c$). In our experiment, $I_c$ ranges from 10 to 49 μA for five different devices at $T = 260$ mK, corresponding to a critical current density of $J_c = (1.0 - 3.3) \times 10^5$ A/cm$^2$, which is about 100 times larger than that of semiconductor NW junctions.[13, 14]

Figure 2(b) is the dynamic conductance – voltage ($dI/dV$ – $V$) curve of the sample **D1** at 260 mK. The dynamic conductance was measured using an AC lock-in technique. Clearly shown in Fig. 2(b) are sub-gap structures (SGSs) in the dynamic conductance. The conductance peaks are well fitted to the formula $V_n = 2\Delta_{BCS}/ne$ [Fig. 2(c)] for integer $n$ up to 17 ($e$ is the electron charge), from which the superconducting gap $\Delta_{BCS}$ of Al is estimated to be $173 \pm 8$ μeV. This is strong evidence of MARs.[3] Within this picture, a quasiparticle undergoes $n$-1 consecutive Andreev reflections in an SNS junction at finite bias $V_n \leq V < V_{n+1}$ before it escapes the N region with an energy exceeding $2\Delta_{BCS}$.[23] Although the SGSs in the dynamic conductance curve have previously been observed in mesoscopic Josephson junctions of various nanostructures,[10, 11, 13-17, 24] they have rarely been observed in *metallic* SNS junctions.[12] To the best of our knowledge, this is the first clear observation of the SGSs arising from MARs in an all-metallic SNS junction.

The evolution of dynamic conductance curves with temperature is shown in Fig. 3(a), where multiple peaks persist even near $T_c$. This robustness of MARs against temperature is in striking contrast to the exponentially decaying Josephson coupling, as revealed in the $I_c(T)$ curve. Since each peak value $V_n$ is scaled with $\Delta_{BCS}$, the temperature dependence of $\Delta_{BCS}$ can be extracted from $V_n(T)$. Shown in Fig. 3(b) are measured peak values $V_n$ with $n = 2, 3$, and 4 at different temperatures, which are in good agreement with the Bardeen-Cooper-Schrieffer (BCS) theory of superconductivity.[25]

A highly transparent NS interface is required to explain our observation of multiple peaks in the differential conductance curves. For our device with a single-crystalline Au NW, the transparency of the NS interface surpasses that of conventional NS interfaces fabricated by a top-down approach such as the shadow evaporation technique.[18] The interface transparency ($T_{int}$) can be estimated from the excess current ($I_{exc}$),[26] which is the current at the crossing point of the extrapolated line of the $I$–$V$ curve from the high bias ($eV > 2\Delta_{BCS}$) region to the zero $V$-axis.[27] In our experiment, $eI_{exc}R_n/\Delta_{BCS}$ ranges from 0.36 to 1.98, resulting in $T_{int} = 0.55 - 0.95$, which is among the highest values observed in superconducting proximity junctions.[14] Such high transparency is attributed to the single crystalline Au NW with a chemically inert surface. It is also worth noting that $T_{int}$ is proportional to the $I_cR_n$ product in our experimental range [see Fig. 3(c)].

One of the unique features observed in our devices is the occurrence of hysteresis at both low- (near $I_c$) and high- ($> I_c$) bias currents. We first discuss the lower-bias hysteresis. When the current is swept down, reversed switching from resistive to supercurrent branch occurs at a retrapping current $I_r$ ($<I_c$), giving a low-bias hysteretic to the $I$–$V$ curve. The hysteresis is predominant at temperatures lower than 0.5 K, as shown in Fig. 3(d). Similar behaviour has been observed in the weak links of semiconductor NWs,[13, 14] carbon nanotubes,[15, 16] graphene,[17] and normal metals.[28-30] There have been several suggestions for the possible origin of such low-bias hysteresis: (i) the resistively- and capacitively-shunted junction (RCSJ) model,[25] (ii) a self-heating effect,[31] and (iii) conductance enhancement due to MARs.[4]

The simple RCSJ model is not suitable for the present observations, since we are utilizing the inert noble metal Au for the weak link, the geometrical junction capacitance of which is almost negligible. Instead, the effective capacitance ($C_{eff}$) can be defined from the electron diffusion time in an Au NW ($\tau = L^2/D$, where $D$ is the electron diffusion constant),[29] resulting in $C_{eff} = \tau R_n^{-1} \sim 10$ pF. Consequently, the quality factor of $I_c/I_r = (2eI_cR_n\tau/\hbar)^{1/2}$, where $\hbar$ is the Planck constant, is estimated to be about 2.8. This is comparable to the observed value.

A self-heating effect may be induced by the Joule power. The power density just above $I_c$ is roughly 230 nW/μm,[3] which is at least two orders of magnitude larger than the typical value necessary for electron heating,[31] implying that the Joule heating is not negligible. However, considering the high thermal conductivity of Au NW, it cannot be conclusively stated that a self-heating effect is the primary origin of the low-bias hysteresis in our system.

Finally, the conductance enhancement due to the MARs can give rise to hysteresis in the $I$-$V$ curve.[4] If the transparency of the NS interface is adequate and the inelastic mean-free path (phase breaking length, $L_\varphi$) is sufficiently larger than $L$, then each Andreev reflection process gives rise to an increase of the conductance and, consequently, low-bias hysteresis. Though plausible, there is no direct evidence of this scenario yet.

Another interesting feature of a long diffusive SNS junction is the temperature dependence of the critical current, $I_c(T)$, shown in Fig. 3(d). We have fitted $I_c(T)$ to the theoretical results[32] of $eI_cR_n = aE_{Th}[1-b\exp(-aE_{Th}/3.2k_BT)]$, where $E_{Th}$ is the Thouless energy and $k_B$ the Boltzmann constant. The best-fit is obtained with $E_{Th}$ = 16 μeV, $a$ = 6.7 and $b$ = 1.4. Given the values $a$ = 10.82 and $b$ = 1.30 in the extreme long junction limit, the fitted results are quite reasonable. In particular, the fitted Thouless energy $E_{Th}$ is compared with the value $E_{Th,exp}$ obtained from the relation $E_{Th,exp} = \hbar D/L^2$ with $D = v_F l_e/3$, where $v_F$ is the Fermi velocity and $l_e$ is the elastic mean free path of Au. By using $v_F$ = 1.39 × 10$^8$ cm/s and the resistivity of Au NW $\rho_{Au}$ = 4.3 μΩcm, we obtain $l_e$ = 19 nm, $D$ = 88 cm$^2$/sec, and $E_{Th,exp}$ = 19 μeV, which is quite close to the fitted result of $E_{Th}$. In addition, the requirements of a long and diffusive junction are also satisfied with $L > \xi \gg l_e$, where $\xi = (\hbar D/\Delta_{BCS})^{1/2}$ = 180 nm is the superconducting coherence length.

Here, we emphasize that the $I_cR_n$ product, a figure of merit for the Josephson junction, is comparable to $\Delta_{BCS}/e$. The inset of Fig. 3(a) displays $eI_cR_n/\Delta_{BCS}$ = 0.4 − 1.5 obtained at $T$ = 260 mK from five different devices. This high value of $I_cR_n$ is quite striking, as $eI_cR_n/\Delta_{BCS}$<0.1 in previous works on SNS junctions fabricated by the shadow evaporation technique, even at lower temperatures.[29, 30] Theoretically, $eI_cR_n$ = 10.8 $E_{Th}$ was predicted at zero temperature in the long-junction limit ($E_{Th}/\Delta_{BCS} \ll 1$),[32] and $eI_cR_n$ = 2.08 $\Delta_{BCS}$ in the short-junction regime ($E_{Th}/\Delta_{BCS} \gg 1$).[33] In the present experiment, $E_{Th}/\Delta_{BCS}$ ranges from 0.02 to 0.11, which indicates that our Au NW junctions belong to a crossover regime between two extreme limits and thus give rise to relatively high $I_cR_n$ values.

To gain further insight into the Au NW SNS junctions, we investigated the effect of an external microwave field. According to the AC Josephson effect,[25] oscillating supercurrent with a voltage-dependent Josephson frequency $f_J = 2eV/h$ can be synchronized to applied microwave frequency $f_{mw}$. The *I–V* curve of the SNS junction is thereupon expected to exhibit quantized voltage plateaus (so-called Shapiro steps) at $V_m = mhf_{mw}/2e$ ($m$ is an integer). A representative *I–V* curve for the Au NW junctions is shown in Fig. 4 for $f_{mw}$ = 4.6 GHz at $T$ = 260 mK. Voltage plateaus are clearly visible and regularly spaced on the *V*-axis, but the step order (*m*) turns out to take not only integer but also half-integer values. For the integer Shapiro steps, a linear relationship between the voltage spacing $\delta V = V_m - V_{m-1}$ and $f_{mw}$ is verified for $f_{mw}$ = 2 – 8 GHz in the inset, and the proportional coefficient, $\delta V/f_{mw}$, is in good agreement with the quantum constant of $h/2e$ = 2.07 µV/GHz.

In summary, we have fabricated mesoscopic Josephson junctions using single crystalline Au NWs contacted with superconducting Al electrodes. A highly pronounced $I_cR_n$ product and high-order (*n* = 17) sub-gap structures of *dI*/*dV* imply that the coherent MARs are the dominant electronic transport mechanism over the whole system, which differs significantly from previous studies on conventional SNS junctions. We anticipate that the single crystalline Au nanowires will provide an ideal platform for SNS mesoscopic Josephson junctions for future applications involving quantum information devices.

**Methods**

Au NWs were synthesized on a c-cut sapphire substrate in a horizontal quartz tube furnace system by using a vapor transport method. The sapphire substrate was placed a few cm downstream from an alumina boat that was filled with 0.03 g of pure Au powder as a precursor. The Ar gas flowed at a rate of 100 sccm, keeping the chamber pressure at 1 ~ 5 Torr. The high-temperature zone of the furnace was heated up to 1100 °C. A typical reaction time for the NW growth was about 30 min. The single crystalline Au NWs have a diamond-shaped cross section with a diameter of 80 ~ 130 nm and a length of tens of micrometers. For the fabrication of the Au NW junctions, a droplet of the solution containing Au NWs was deposited on a n$^+$ Si substrate with a 300 nm-thick $SiO_2$ top layer. After defining electrode patterns by standard electron-beam lithography, the metal electrodes were formed by depositing 200 nm-thick Al layers in an ultra-high vacuum chamber. Here the Al electrode was used as a superconducting contact. A representative SEM image and a schematic view of the device are shown in Fig. 1(b) and the inset, respectively. The normal-state resistance ($R_n$) of the Au NW junctions ranges from 4 to 10 Ω, from which the upper limit of the Au NW resistivity can be estimated to be $\rho_{Au}$ = 14 ± 8 μΩcm at $T$ = 2 K.

**Acknowledgements**

Dr. M. Jung thanks Prof. Kazuhiko Hirakawa for useful discussions. This work was supported by MEST through NRF (2008-2002795, 2009-0080453) and 'Center for Nanostructured Material Technology' under '21 C Frontier R&D Programs' (2009K000468) of MEST, Korea.

Correspondence to jinhee@kriss.re.kr or bongsoo@kaist.ac.kr.

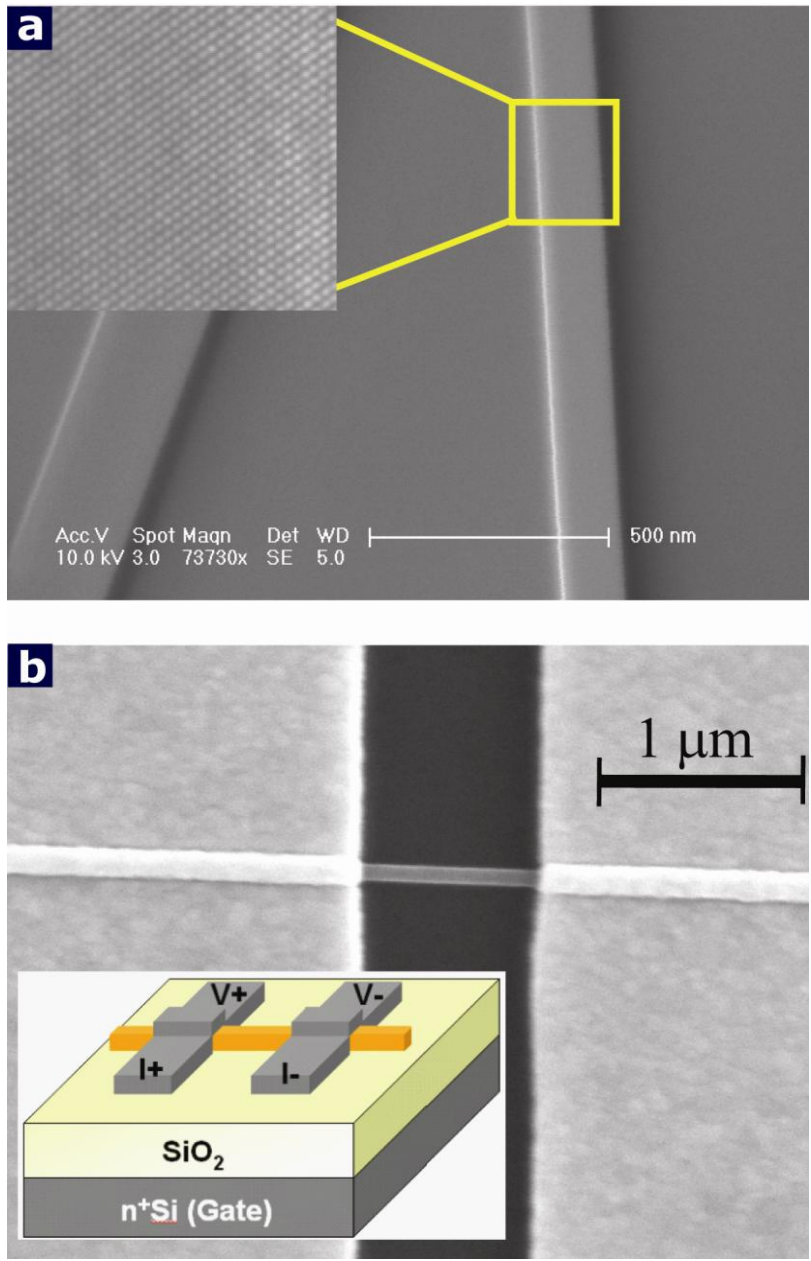

**Figure 1.** (a**)** Scanning electron micrograph and a high-resolution TEM image (inset) of single crystalline Au NWs. (b**)** SEM image of the Au NW device. Inset: a schematic view of the device and the measurement configuration.

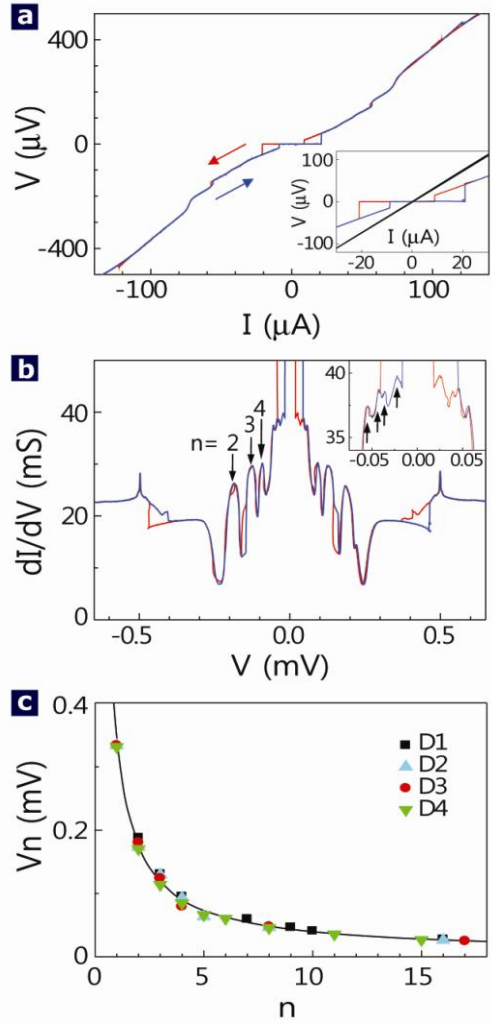

**Figure 2.** (a) Current–voltage (*I–V*) characteristics of device **D1** at *T* = 260 mK. The blue and red lines indicate increasing and decreasing current direction, respectively. Inset: magnified view of the *I–V* curve near zero bias. The normal-state *I–V* curve (black line) was obtained at the same *T* with application of magnetic field *H* = 0.1 T. (b) Dynamic conductance - voltage (d*I*/d*V*-*V*) characteristics for **D1** at *T* = 260 mK with increasing (blue) and decreasing (red) current, respectively. The d*I*/d*V* peaks at $V_n = 2\Delta_{BCS}/ne$ (*n* = 2, 3, 4) are denoted as arrows. Inset: enlarged view of the d*I*/d*V*–*V* curve around zero bias. High-order sub-gap peaks are indicated by arrows (*n* = 7, 9, 10, 16 from left to right). (b) The peak voltage ($V_n$) versus the index *n* for four different devices. The solid line is fitted to $V_n = 2\Delta_{BCS}/ne$ with $\Delta_{BCS}$ = 173 μeV.

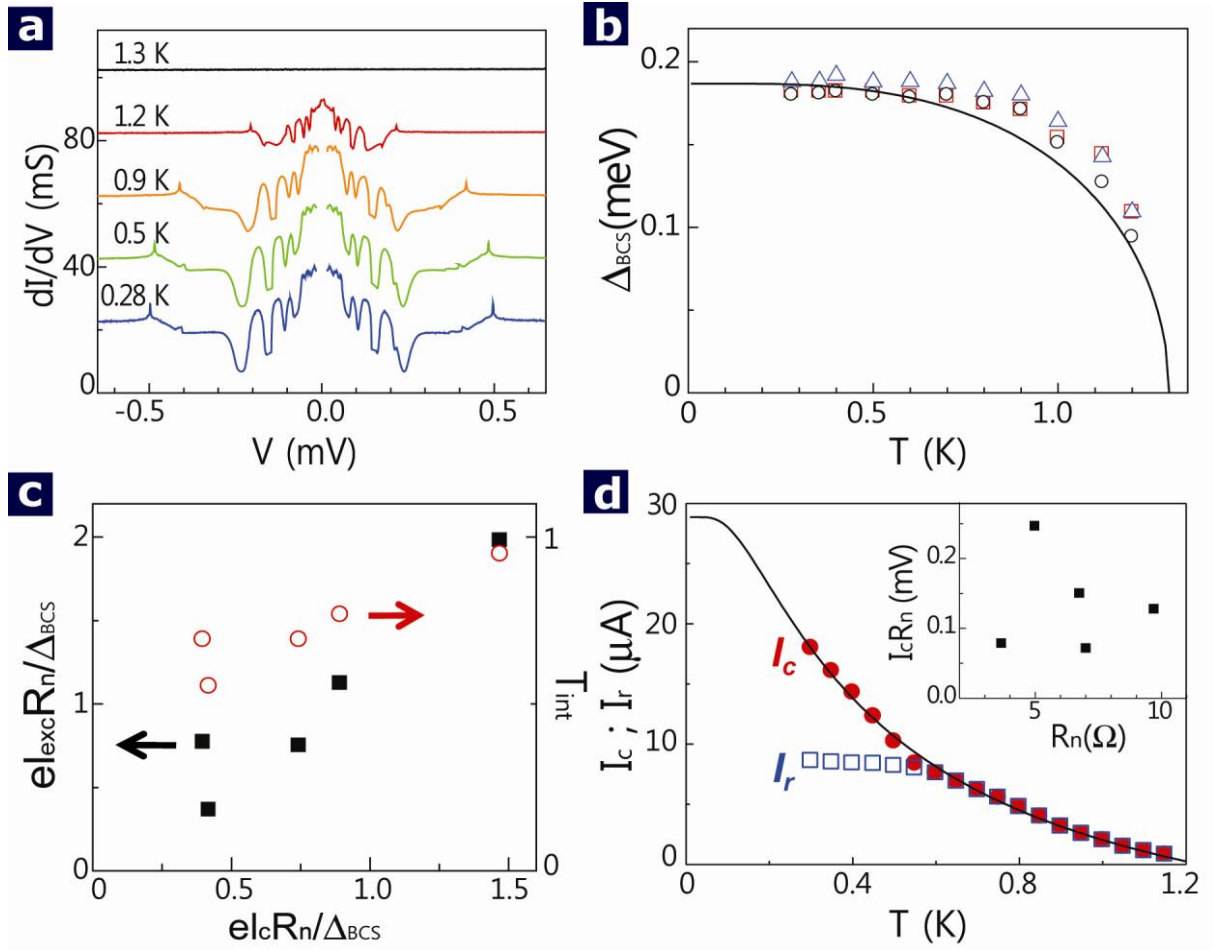

**Figure 3.** (a) Evolution of *dI/dV–V* curves of sample **D1** with temperature. The measured temperatures are 0.28, 0.50, 0.90, 1.2, and 1.3 K from bottom to top. (b) Temperature dependence of $\Delta_{BCS}$ estimated from the sub-gap structures with *n* = 2 (square), 3 (triangle), and 4 (circle). The solid line is fitted to $\Delta_{BCS}(T)$ predicted in BCS theory. (c) The excess current ($I_{exec}$; square) and the transparency ($T_{int}$; circle) depending on the reduced $I_cR_n$ product for five different samples. (d) Temperature dependence of critical ($I_c$; circle) and retrapping ($I_r$; square) currents. The solid line is a theoretical fit of $I_c$, as explained in the text. Inset: the $I_cR_n$ product and the normal-state resistance $R_n$ for five different devices. $I_c$ was obtained at a base temperature of *T* = 260 mK.

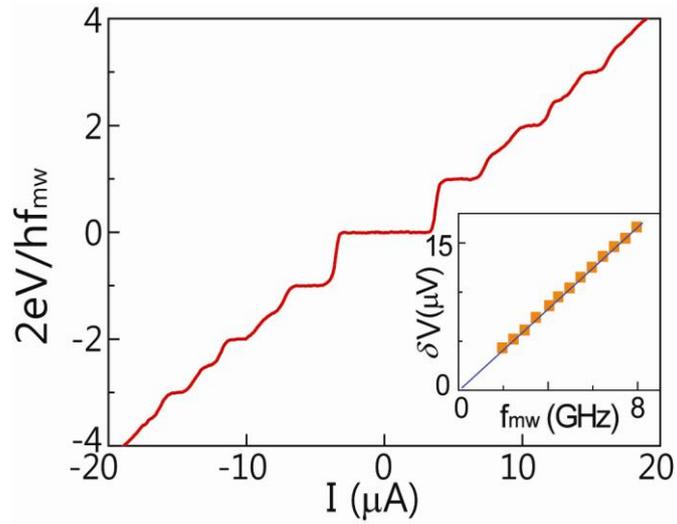

**Figure 4.** *I–V* characteristics under a microwave field of frequency $f_{mw}$ = 4.6 GHz at $T$ = 260 mK. *V* was normalized by $hf_{mw}/2e$ to make the integer and the half-integer Shapiro steps clearly identifiable. Inset: voltage spacing unit ($\delta V$) between the integer steps versus $f_{mw}$. The solid line is a direct plot of $hf_{mw}/2e$ without any adjustable parameters.

16